\documentclass[conference]{IEEEtran}
\usepackage{cite}
\usepackage{amsmath,amssymb}
\usepackage{graphicx,booktabs,placeins,url}
\newif\iffourfigures\fourfigurestrue
\begin{document}
\title{From Connection Loss to Collision: Closed-Loop Fault Assessment of a Fly-Connectome SNN}
\author{\IEEEauthorblockN{Shuchao Gao}
\IEEEauthorblockA{\textit{Graduate School of Information, Production and Systems, Waseda University, Kitakyushu, Japan}\\
gaoshuchao@fuji.waseda.jp}}
\maketitle
\bstctlcite{IEEEcontrol}
\begin{abstract}
The public release of the complete adult fruit-fly brain connectome enables brain-scale spiking-network studies, motivating closed-loop assessment of connection loss. We test a connectome-derived controller in synthetic pursuit--escape and check record consistency. Across 576 test episodes in paired campaigns, the baseline mapping at 50\% connection-row deletion yields 11/48 proportional and 0/48 relay collisions by 1\,s, despite paired P/R trigger/burst schedules. The prespecified secondary relay-minus-proportional net-loss difference is $-22.9$ percentage points (95\% seed-cluster CI $[-33.3,-12.5]$). Two separately recalibrated mappings yield smaller, uncertain point estimates; unequal healthy clearance, recalibration and different test cohorts limit attribution. A frozen command-rate warning detects only 2/13 baseline collisions across all tested deletion levels. In a separate author-designed fly/delayed-tank catalogue, checks plus model-specific replay detect 81/87 corruptions versus 36/87 for trusted headers, accepting all 20 distinct valid inputs (six misses). These results quantify finite-horizon collisions and the gains and limits of record checks.

\end{abstract}
\newcommand{\SystemFigure}{%
\begin{figure*}[t]
\centering
\includegraphics[width=\textwidth]{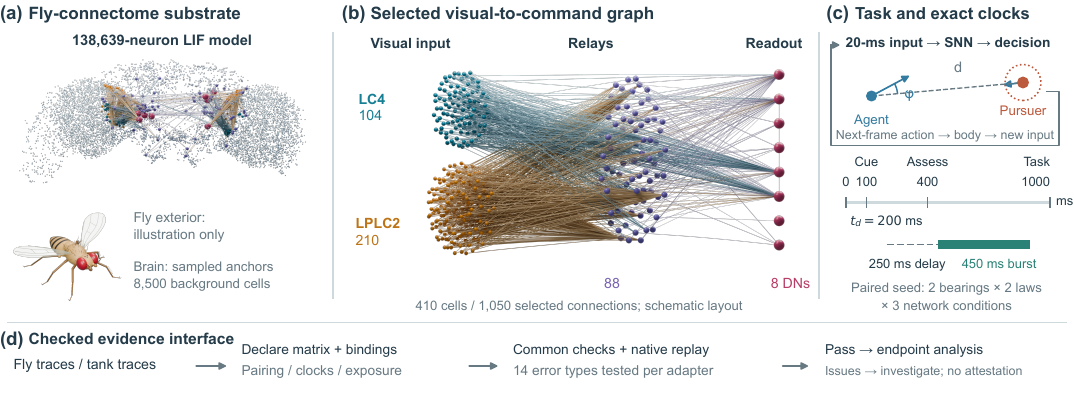}
\caption{System and checked assessment interface. (a) Model anchors and an illustrative fly; (b) schematic 410-cell/1,050-connection graph, not measured neurites or the full recurrent network. (c) Sampled feedback and exact burst events; $t_d=200$\,ms is illustrative. (d) Common checks and native replay; the 14 types are development challenges, separate from Fig.~\ref{fig:v10f-reuse}'s catalogue.}
\label{fig:v10f-system}
\end{figure*}}
\newcommand{\ReuseCaption}{Cross-plant example and frozen-checker ablation. (a) First tank seed, high inflow, proportional law; zoomed level axis, safe interval $[0.2,0.8]$; all 96 tasks remain safe. (b) Requested outflow at 0.60\,s applies at 0.75\,s on 100-ms frames. (c) Invalid cases detected: A headers, B=A+common, C=A+native, D=full. Fly/tank invalid denominators are 45/42, with 12 valid controls each. Designed cases are not general coverage.}
\newcommand{\WarningCaption}{Frozen secondary warning operating points. Rows state contact/success denominators; RF-0/A/B contain 24/12/12 independent seed clusters. Markers show (a) missed-contact and (b) false-alarm rates with pointwise 95\% whole-seed bootstrap CIs. Activity is two-sided; the forecast holds the current command using exact simulated state. Positive-class support is 12/2/1 seed clusters. Zero-width and class-conditional empirical intervals do not establish zero population risk.}
\newcommand{\reuseRef}{Fig.~\ref{fig:v10f-reuse}}
\newcommand{\warningRef}{Fig.~\ref{fig:v10f-warning}}
\newcommand{\ReuseFigure}{%
\begin{figure}[t]\centering
\includegraphics[width=\columnwidth]{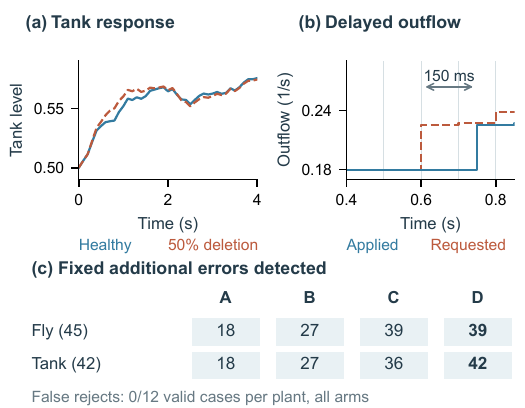}
\caption{\ReuseCaption}\label{fig:v10f-reuse}
\end{figure}}
\newcommand{\WarningFigure}{%
\begin{figure}[t]\centering
\includegraphics[width=\columnwidth]{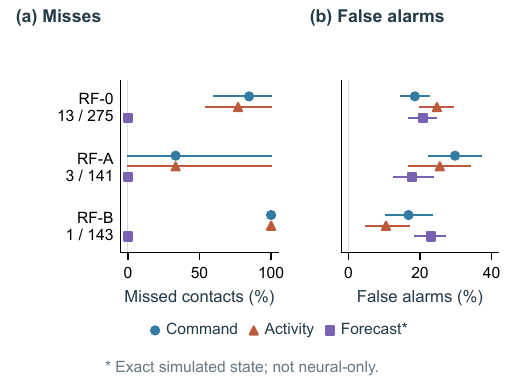}
\caption{\WarningCaption}\label{fig:v10f-warning}
\end{figure}}
\newcommand{\TransferFigure}{%
\begin{figure*}[t]\centering
\includegraphics[width=\textwidth]{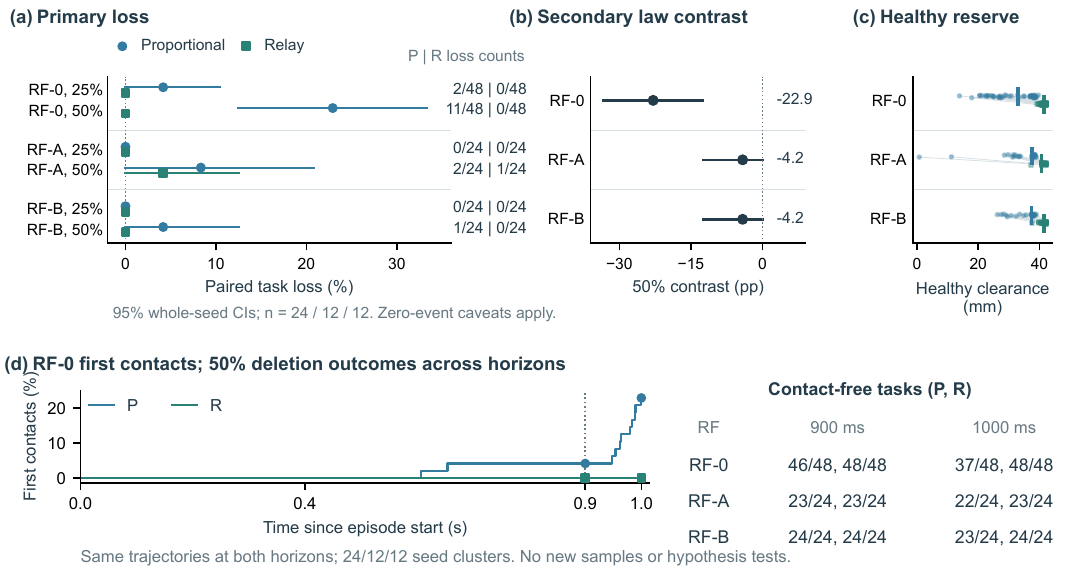}
\caption{Paired loss across sensory realizations and horizons. (a) Primary losses and (b) prespecified secondary R-minus-P contrasts (percentage points), with pointwise 95\% whole-seed CIs, $n=24/12/12$, conditional on saved calibrations. All healthy tasks succeed; no gains occur. (c) Paired healthy minimum-clearance points and median marks show unequal reserve. (d) RF-0/50\% first contacts and all-map 900/1000-ms success counts (P, R); both endpoints reuse the same 24/12/12 seed clusters, not new replicates or tests. Maps differ in draws, recalibration and cohorts, not an isolated RF effect.}
\label{fig:v10f-transfer}
\end{figure*}}

\section{Introduction and Positioning}
How does matched connection loss in a connectome-derived spiking neural network (SNN) affect first contact as feedback, observation horizon and calibrated sensing change? We pair healthy and damaged runs, resolving neural commands, action timing and contact separately.

Shiu's LIF model\cite{shiu2024drosophila}, derived from the publicly released adult-fly connectome\cite{dorkenwald2024neuronal}, supplies commands to our synthetic pursuit--escape plant. An\cite{an2026sparse} studies task/state-dependent sparsification; Jo\cite{jo2026simulating} links damaged circuits to virtual-arena behavior. We quantify paired first contact under frozen laws and horizons, and test future-contact warnings from a fixed 400-ms prefix.

DriveFI searches closed-loop safety violations\cite{jha2019drivefi}; SpikeFI, SpikingJET and enpheeph provide SNN fault injection\cite{spyrou2024spikefi,gogebakan2024spikingjet,colucci2022enpheeph}. Common random numbers\cite{glasserman1992crn}, trajectory monitoring\cite{donze2010breach}, benchmarking\cite{yik2023neurobench,dangelo2026embodied} and physical fault propagation\cite{chessa2026crosslayer} are established. Our ablation tests which pairing, timing and native-replay obligations add findings beyond trusted record identities.

We contribute paired, stage-resolved task measurements of law-discordant contacts under shared schedules, horizon sensitivity, uncertain RF transfer and frozen-prefix warning failures. A separate frozen-checker ablation quantifies added consistency findings, costs and misses across fly and tank adapters. Together these measure conditional task preservation and record-consistency audit gains; timing remains descriptive.

\SystemFigure
\ReuseFigure
\TransferFigure
\section{Neural Experiment and Statistical Design}
\subsection{Network, Feedback and Calibration}
The FlyWire v783-derived model\cite{dorkenwald2024neuronal} has 138,639 LIF neurons and 15,091,983 signed, contact-weighted connection rows; cell labels use FlyWire annotations\cite{schlegel2024annotation}. Shiu parameters\cite{shiu2024drosophila} run in Brian~2\cite{stimberg2019brian2}, with 0.1-ms steps, 2.2-ms sensory refractoriness and state/weight restoration per episode. Assumed RFs stimulate 104 LC4 and 210 LPLC2 cells; eight descending cells provide command features (Fig.~\ref{fig:v10f-system}(a--c)).

Between events, $\dot v=(v_0-v+g)/(20\,\mathrm{ms})$ and $\dot g=-g/(5\,\mathrm{ms})$. At $-45$\,mV, reset $(v,g)=(v_0,0)$, $v_0=-52$\,mV; both derivatives pause during refractoriness. Recurrent events add 0.275\,mV per signed contact to $g$ after 1.8\,ms; external events add 68.75\,mV to $v$.

Input starts at 100\,ms with visual period $\delta_{\rm v}=0.020$\,s. In degrees, $\beta$ is world bearing minus heading, $a=(180/\pi)\arctan(0.05/d)$ is half-angle at separation $d$, $\sigma=\max(30^\circ,a)$ and $\ell=(a-a_{\rm prev})/\delta_{\rm v}$:
\begin{equation}
\begin{aligned}
u&=\operatorname{clip}_{[0,1]}[\ell_+e^{-a/60}/150],\\
\lambda_i&=\operatorname{clip}_{[0,150]}[150w_i u e^{-[\operatorname{wrap}(\phi_i-\beta)^2+\psi_i^2]/(2\sigma^2)}]\ \mathrm{Hz}.
\end{aligned}
\end{equation}
Pre-cue $u=0$ masks the initial analytic expansion; the 100-ms input uses 80--100-ms angle history. Wrap maps to $[-180^\circ,180^\circ)$. RF azimuth/elevation $\phi_i,\psi_i$ set $w_i=1$ for LC4 and $1+0.8(\psi_i+30)/90$ for LPLC2; other external rates are zero. A 0.3-m/s pursuer starts 0.25\,m away at $\pm30^\circ$ and targets the agent's segment-start position. Linear segments use starting headings; agent heading updates at segment end, capped at $\pm100^\circ$. Decisions apply next frame. First trigger starts a heading-fixed 0.35-m/s burst 250\,ms later for 450\,ms; otherwise speed is 0.05\,m/s. Actuator events split segments; first contact is solved within them at 0.05-m center separation. Failure latches; resources continue to $H=1$\,s; warnings use pre-contact data.

Per map, 20 passive development episodes (four seeds, five bearings in $[-60^\circ,60^\circ]$) fit affine score $z$. Standardized eight-cell five-frame means at frame ends $200\le t\le400$\,ms predict bearing/$60^\circ$ with ridge penalty 10 and an unpenalized intercept. Trigger threshold is half the development tenth percentile of 200--300-ms episode command means, floored at 2.5\,Hz. Before cue or below threshold, steering is zero; otherwise P/R apply $-600\operatorname{clip}(z,-1,1)$/$-600\operatorname{sgn}(z)$ deg/s outside the burst. Eight held-out validation clusters per RF give RMSE $65.9^\circ/67.8^\circ/67.2^\circ$: these measure continuous-angle error, not directional sign accuracy.

\subsection{Pairing, Cohorts and Endpoints}
Shared uniforms give nested deletions $w_j^{(f)}=w_j\mathbf1[V_{sj}\geq f]$ and input events $\mathbf1[U_{sk\tau i}<\lambda_{ki}\Delta t]$. Indices denote seed $s$, frame $k$, neural step $\tau$, cell $i$ and connection row $j$; $\Delta t=10^{-4}$\,s. Feedback changes $\lambda$, while input variates and row masks remain paired. Seeds and hashes are traceable, without cross-platform bit-identity guarantees.

RF-0/A/B use 24/12/12 independent test clusters, each containing both bearings, two laws and healthy/25\%/50\% deletion: 288/144/144 episodes. New maps independently redraw 314 azimuths from the same hemisphere prior (left $10$--$90^\circ$, right $-90$--$-10^\circ$), preserving identities, elevation and gradients. Each repeats the frozen development recipe, validates on 32 healthy episodes in eight separate clusters, then freezes thresholds. All maps, seeds and failures remain; no test tuning or efficacy stopping occurs. Transfer includes recalibration, not zero-shot or RF-isolated causality.

With $Y=1$ meaning no contact through $H$, registered primary endpoints are paired loss $L=Y_H(1-Y_F)$, gain $G=(1-Y_H)Y_F$ and activity exposure. Net loss $D=L-G$ is derived. The relay-minus-proportional contrast and warnings are prespecified secondary; trigger-timing, stage and clearance/heading diagnostics are post-hoc. For cumulative spikes $S$ and $T_*=\min(T_F,T_H)$, own-stop and common-duration ratios are
\begin{equation}
R_{\rm stop}=\frac{S_F(T_F)}{S_H(T_H)},\qquad
R_{\rm common}=\frac{S_F(T_*)}{S_H(T_*)}.
\end{equation}
Here $T$ is contact-frame end or $H$, at 20-ms resolution. Common duration remains outcome-dependent and does not equate sensory histories; fixed 400-ms/1-s windows are separate endpoints.

Binary loss/gain average both bearings within seed, then seeds. Exposure uses the median of paired ratios, retaining both bearings; warnings pool trial-level confusion counts. Pointwise 95\% CIs resample whole seeds 20,000 times (RNG 20261022 for RF-0; 20261023 for transfer), retaining their complete condition matrix and recomputing the corresponding statistic. CIs condition on stored maps/calibrations. Symmetric-null sign-flip tests use separate Holm families for two baseline and four transfer contrasts. Maps and separate angle/action cohorts are not pooled.

Warnings use only the 400-ms prefix. Command count $C$ sums eight cells across fifteen 20-ms bins in 100--400\,ms; saved rates are exact 50-Hz multiples, allowing lossless integer recovery. Each 32-record healthy validation set gives a linear/type-7 fifth percentile: 14/14.55/13.55 counts in RF-0/A/B. The frozen strict rule $C<Q_{.05}$ does not flag ties; displayed $C/2.4$\,Hz never determines the label. Activity uses a law/bearing-specific healthy reference and a frozen 95th-percentile absolute-log-ratio cutoff; an upper-only comparator flags ratios above two. Forecast uses exact simulated state and pending actuation, holding the current turn; if untriggered at 400\,ms it assumes a 420-ms trigger. Its schedule-only baseline sets future turn to zero. Each RF freezes its fifth-percentile forecast cutoff on healthy validation; these forecasts are not neural-only detectors.

Transfer worker-lifetime peaks were 1.20--1.21\,GiB on Xeon W-2223/31.6\,GiB RAM. Elapsed times for 576 test episodes sum to 29.55\,h (median 167\,s), not CPU-hours or campaign duration.

\section{Executable Evidence Contract}
Adapters normalize native traces, binding campaign identities, clocks, resource units/histories, sources, maps, calibration, input/mask digests and first failure (Fig.~\ref{fig:v10f-system}(d)). Shared variates need not yield equal feedback-dependent event hashes; prefixes exclude future labels.

The second plant obeys $\dot h=I-u-0.04h$, $h(0)=0.5$, safe interval $[0.2,0.8]$. Eight seeds, two inflows, two laws and three sensor-deletion levels give 96 safe examples; missing samples hold their last value (\reuseRef(a)). Native replay checks analytic transitions and 150-ms actuation on a 100-ms grid (\reuseRef(b)).

Before scoring, we froze 111 software case IDs: 87 corruptions and 24 valid IDs (107 distinct inputs; 20 distinct valid inputs), not independent statistical samples. Valid controls retain off-grid actuation, worker partitions and feedback-dependent histories. This author-designed catalogue is not blind generalization evidence.

Arm A checks schema, matrix membership and trusted input/configuration/mask headers, not whole-record hashes. B adds unchanged common pairing, clock, exposure, cumulative-resource and label invariants; C adds native input/actuator/plant and seeded-mask replay; D combines them. All arms use identical normalized rows without short-circuiting; registry setup is timed separately.

\section{Results and Conditional Interpretation}
\subsection{First-Contact Loss Depends on the Declared Horizon}
Fig.~\ref{fig:v10f-transfer}(a) reports paired losses; Fig.~\ref{fig:v10f-transfer}(d) rescores the same 50\%-deletion trajectories at 900/1000\,ms across all RFs. RF-0 has nine additional P contacts in the final 100\,ms (latest 999.962\,ms); law-discordant RF-A/B contacts also occur after 900\,ms. These descriptive rescorings add neither samples nor confirmatory tests; RF-0's repeated endpoints were registered.

The prespecified secondary R-minus-P net-loss contrast (Fig.~\ref{fig:v10f-transfer}(b)) was $-22.9$\,pp (95\% CI $[-33.3,-12.5]$; Holm $p=0.00195$). A/B each yielded $-4.2$\,pp, CI $[-12.5,0]$ (all four transfer Holm $p=1$; one nonzero seed per 50\% contrast). This establishes neither equivalence nor an isolated RF effect; no between-map test was run. RF-0/25\% remains uncertain ($p=0.50$). A/B have no 25\% loss; twelve event-free clusters permit a one-sided 95\% upper bound of 22.1\% for any loss in a new two-bearing cluster.

All 576 records triggered. Post-hoc fault-minus-healthy trigger-time differences at 50\% deletion averaged 80.0/57.5/57.5\,ms in RF-0/A/B (SD 42.9/46.6/29.7\,ms across 48/24/24 bearing pairs; 24/12/12 independent seeds, P/R duplicates counted once). Trigger means first post-cue crossing by the eight-cell five-frame mean at 20-ms resolution. All 288 P/R pairs shared trigger/burst schedules under pre-trigger gating, yet 15 had different contact outcomes. Both RF-0 25\% P contacts occurred without delay: timing alone neither determines outcome across laws nor accounts for every contact.

Among 543 records with no contact by burst completion and positive follow-up, all 3,114 decisions at/after burst end and before contact/$H$ had zero eight-cell features and gated turn; 474 retained whole-network spikes. Thirteen later contacts fall in this observed walking regime, not a universal silence rule or isolated delay mechanism. Exclusions are four prior contacts, 17 ongoing bursts and 12 zero-follow-up records.

Fig.~\ref{fig:v10f-transfer}(c) shows paired healthy minimum-clearance points and vertical median marks: RF-0 P/R medians 33.1/41.4\,mm; 400-ms absolute headings $88.3^\circ/100^\circ$. The contrast compares complete controllers with unequal healthy reserve. Burst-end state and follow-up are possible contributors; RF calibration and cohorts vary together, precluding isolated mechanisms.

\subsection{Frozen Scalar Warnings Miss Subsequent Contacts}
\WarningFigure
Across healthy/25\%/50\%, command warnings detect 2/13, 2/3 and 0/1 contacts (RF-0/A/B; \warningRef(a)); false alarms are 51/275, 42/141 and 24/143 (\warningRef(b)). Exact-count labels/CIs are unchanged; alternative rate arithmetic alerts on 19 ties ($C=14$). Upper-only activity detects none. Exact-state forecasts detect all contacts but falsely flag 57/275, 25/141 and 33/143 successes; schedule-only flags 84/275. Extra geometry/actuation information precludes neural-only comparison.

Sensitivities rely on 12/2/1 failure clusters. Of 20,000 resamples, 2,217/7,099 lack positives in A/B and are omitted only for the corresponding conditional rate; zero-width intervals imply no zero-risk guarantee. Own-stop/common-duration medians differ by less than 0.006. P/50\% median 400-ms/1-s activity ratios are 0.981/1.505, 0.961/1.212 and 0.956/1.139. A separate eight-cluster $\pm45^\circ$ test flags 22/32 healthy activity records with uncertain law contrast. These results bound the frozen criteria, not all monitoring.

\subsection{Audit Gain and Reporting Impact}
For 87 corruptions, A/B/C/D detected 36/54/75/81 (\reuseRef(c)), rejecting 0/24 valid IDs (20 distinct inputs). Headers missed actuation and first-contact inconsistencies detected by native replay. Above A, six cases were common-only, 27 native-only and 12 shared. Fly C=D=39/45; Full's six additions over Native were all tank label/exposure violations. Full missed three consistent global-count offsets and three altered burst declarations.

Median D/A times per valid fixture were 1.32/0.0017\,s (fly), 0.0168/0.0018\,s (tank); registry setup cost 2.24\,s separately. Process RSS peaked at 446\,MiB with fixtures; a synthetic zero-neural logger passes replay.

In 72 historical V8 records, 17 onsets were 10\,ms late; ten complete bursts lasted 440\,ms (450\,ms declared; six at $H$). Rescoring action reports corrects burst-active counts from 13/17 to 7/17 among records reaching $H$; excess nominal burst exposure totals 180\,ms over all 72 records. Task labels agree with logged geometry; contact or energy effects of corrected actuation cannot be inferred.

\section{Scope, Artifacts and Conclusion}
Evidence concerns one graph, an assumed RF family, a synthetic plant, unequal healthy reserve and sparse failures. Roughly $66^\circ$ held-out RMSE limits bearing reconstruction. Walk-only contact by 0.8\,s reflects plant speeds; one burst followed by slower walking makes 1-s success finite-horizon preservation, not sustained avoidance. Functional row loss establishes no biological tolerance, device upset rate, hardware safety or energy/area benefit; trace consistency cannot authenticate neural execution.

Code and data are fixed\cite{artifactEngineering}; neural reruns require hash-matched upstream matrices.

We quantify conditional task preservation and record-consistency gains. This task-level assessment provides a basis for evaluating how future device-level nonidealities propagate through neuromorphic controllers to closed-loop behavior.

\FloatBarrier\clearpage
\IEEEtriggeratref{9}
\bibliographystyle{IEEEtran}
\bibliography{refs}

\begin{thebibliography}{10}
\providecommand{\url}[1]{#1}
\csname url@samestyle\endcsname
\providecommand{\newblock}{\relax}
\providecommand{\bibinfo}[2]{#2}
\providecommand{\BIBentrySTDinterwordspacing}{\spaceskip=0pt\relax}
\providecommand{\BIBentryALTinterwordstretchfactor}{4}
\providecommand{\BIBentryALTinterwordspacing}{\spaceskip=\fontdimen2\font plus
\BIBentryALTinterwordstretchfactor\fontdimen3\font minus
  \fontdimen4\font\relax}
\providecommand{\BIBforeignlanguage}[2]{{%
\expandafter\ifx\csname l@#1\endcsname\relax
\typeout{** WARNING: IEEEtran.bst: No hyphenation pattern has been}%
\typeout{** loaded for the language `#1'. Using the pattern for}%
\typeout{** the default language instead.}%
\else
\language=\csname l@#1\endcsname
\fi
#2}}
\providecommand{\BIBdecl}{\relax}
\BIBdecl

\bibitem{shiu2024drosophila}
P.~K. Shiu, G.~R. Sterne, N.~Spiller, R.~Franconville, A.~M. Sandoval, J.~Zhou
  \emph{et~al.}, ``A {D}rosophila computational brain model reveals
  sensorimotor processing,'' \emph{Nature}, vol. 634, no. 8032, pp. 210--219,
  2024.

\bibitem{dorkenwald2024neuronal}
S.~Dorkenwald, A.~Matsliah, A.~R. Sterling, P.~Schlegel, S.-C. Yu, C.~E.
  McKellar \emph{et~al.}, ``Neuronal wiring diagram of an adult brain,''
  \emph{Nature}, vol. 634, no. 8032, pp. 124--138, 2024.

\bibitem{an2026sparse}
\BIBentryALTinterwordspacing
D.~An, ``How sparse can a fly-brain model be? {C}onnection strength, wiring
  placement, network state and task jointly determine function-preserving
  compression of the {D}rosophila connectome,'' bioRxiv, 2026. [Online].
  Available: \url{https://www.biorxiv.org/content/10.64898/2026.09.19.752860}
\BIBentrySTDinterwordspacing

\bibitem{jo2026simulating}
\BIBentryALTinterwordspacing
T.~Jo, ``Simulating {A}lzheimer's amyloid-like pathology on a fly central
  nervous system connectome,'' arXiv:2610.19153, 2026. [Online]. Available:
  \url{https://arxiv.org/pdf/2610.19153}
\BIBentrySTDinterwordspacing

\bibitem{jha2019drivefi}
\BIBentryALTinterwordspacing
S.~Jha, S.~S. Banerjee, T.~Tsai, S.~K.~S. Hari, M.~B. Sullivan, Z.~T.
  Kalbarczyk \emph{et~al.}, ``{ML}-based fault injection for autonomous
  vehicles: A case for bayesian fault injection,'' in \emph{2019 49th Annual
  {IEEE/IFIP} International Conference on Dependable Systems and Networks
  ({DSN})}, 2019, pp. 112--124. [Online]. Available:
  \url{https://doi.org/10.1109/DSN.2019.00025}
\BIBentrySTDinterwordspacing

\bibitem{spyrou2024spikefi}
\BIBentryALTinterwordspacing
T.~Spyrou, S.~Hamdioui, and H.-G. Stratigopoulos, ``{SpikeFI}: A fault
  injection framework for spiking neural networks,'' arXiv:2412.06795, 2024.
  [Online]. Available: \url{https://arxiv.org/abs/2412.06795}
\BIBentrySTDinterwordspacing

\bibitem{gogebakan2024spikingjet}
A.~B. G{\"o}{\u{g}}ebakan, E.~Magliano, A.~Carpegna, A.~Ruospo, A.~Savino, and
  S.~{Di Carlo}, ``{S}piking{JET}: Enhancing fault injection for fully and
  convolutional spiking neural networks,'' in \emph{2024 IEEE 30th
  International Symposium on On-Line Testing and Robust System Design
  (IOLTS)}.\hskip 1em plus 0.5em minus 0.4em\relax IEEE, 2024, pp. 1--7.

\bibitem{colucci2022enpheeph}
\BIBentryALTinterwordspacing
A.~Colucci, A.~Steininger, and M.~Shafique, ``enpheeph: A fault injection
  framework for spiking and compressed deep neural networks,'' in \emph{2022
  IEEE/RSJ International Conference on Intelligent Robots and Systems
  (IROS)}.\hskip 1em plus 0.5em minus 0.4em\relax IEEE, Oct. 2022, pp.
  5155--5162. [Online]. Available:
  \url{https://doi.org/10.1109/IROS47612.2022.9982181}
\BIBentrySTDinterwordspacing

\bibitem{glasserman1992crn}
P.~Glasserman and D.~D. Yao, ``Some guidelines and guarantees for common random
  numbers,'' \emph{Management Science}, vol.~38, no.~6, pp. 884--908, 1992.

\bibitem{donze2010breach}
A.~Donz{\'e}, ``Breach, a toolbox for verification and parameter synthesis of
  hybrid systems,'' in \emph{Computer Aided Verification (CAV)}.\hskip 1em plus
  0.5em minus 0.4em\relax Springer, 2010, pp. 167--170.

\bibitem{yik2023neurobench}
\BIBentryALTinterwordspacing
J.~Yik, K.~Van~den Berghe, D.~den Blanken, Y.~Bouhadjar, M.~Fabre, P.~Hueber
  \emph{et~al.}, ``The {NeuroBench} framework for benchmarking neuromorphic
  computing algorithms and systems,'' \emph{Nature Communications}, vol.~16,
  no.~1, p. 1545, Feb. 2025. [Online]. Available:
  \url{https://doi.org/10.1038/s41467-025-56739-4}
\BIBentrySTDinterwordspacing

\bibitem{dangelo2026embodied}
\BIBentryALTinterwordspacing
G.~D'Angelo, J.~E. Pedersen, T.~Hassan, M.~Cianchetti, J.~Bongard, F.~Iida
  \emph{et~al.}, ``A benchmarking framework for embodied neuromorphic agents,''
  \emph{Nature Machine Intelligence}, vol.~8, no.~3, pp. 300--312, Mar. 2026.
  [Online]. Available: \url{http://dx.doi.org/10.1038/s42256-026-01197-w}
\BIBentrySTDinterwordspacing

\bibitem{chessa2026crosslayer}
G.~Chessa, Y.~Yang, E.~Jones, M.~Favalli, and D.~Bertozzi, ``Cross-layer
  analysis of permanent faults in asynchronous {NoCs} for neuromorphic
  computing,'' in \emph{Proceedings of the International Conference on
  Neuromorphic Systems}.\hskip 1em plus 0.5em minus 0.4em\relax ACM, 2026, pp.
  31--38.

\bibitem{schlegel2024annotation}
P.~Schlegel, Y.~Yin, A.~S. Bates, S.~Dorkenwald, K.~Eichler, P.~Brooks
  \emph{et~al.}, ``Whole-brain annotation and multi-connectome cell typing of
  {Drosophila},'' \emph{Nature}, vol. 634, no. 8032, pp. 139--152, 2024.

\bibitem{stimberg2019brian2}
\BIBentryALTinterwordspacing
M.~Stimberg, R.~Brette, and D.~F. Goodman, ``Brian 2, an intuitive and
  efficient neural simulator,'' \emph{eLife}, vol.~8, p. e47314, Aug. 2019.
  [Online]. Available: \url{https://doi.org/10.7554/eLife.47314}
\BIBentrySTDinterwordspacing

\bibitem{artifactEngineering}
\BIBentryALTinterwordspacing
S.~Gao, ``Paired closed-loop fault assessment: Fixed checker ablation and
  historical reporting impact,'' Git commit b29f089, 2026. [Online]. Available:
  \url{https://github.com/Scophield/drosophila-fault-injection/tree/b29f08970f7749c916a472e835517547d832e6cc}
\BIBentrySTDinterwordspacing

\end{thebibliography}
\end{document}